# Congestion Management for Mobility-on-Demand Schemes that use Electric Vehicles

Emmanouil S. Rigas[1] and Konstantinos S. Tsompanidis[2]

[1] Department of Informatics, Aristotle University of Thessaloniki, 54124,
Thessaloniki, Greece
`erigas@csd.auth.gr`
[2] Department of Computing, The University of Northampton, NN15PH,
Northampton, UK
`ktsompanidis@hotmail.com`

**Abstract.** To date the majority of commuters use their privately owned
vehicle that uses an internal combustion engine. This transportation
model suffers from low vehicle utilization and causes environmental pol-
lution. This paper studies the use of Electric Vehicles (EVs) operating in
a Mobility-on-Demand (MoD) scheme and tackles the related manage-
ment challenges. We assume a number of customers acting as cooperative
agents requesting a set of alternative trips and EVs distributed across
a number of pick-up and drop-off stations. In this setting, we propose
congestion management algorithms which take as input the trip requests
and calculate the EV-to-customer assignment aiming to maximize trip
execution by keeping the system balanced in terms of matching demand
and supply. We propose a Mixed-Integer-Programming (MIP) optimal
offline solution which assumes full knowledge of customer demand and
an equivalent online greedy algorithm that can operate in real time.
The online algorithm uses three alternative heuristic functions in decid-
ing whether to execute a customer request: (a) The sum of squares of
all EVs in all stations, (b) the percentage of trips' destination location
fullness and (c) a random choice of trip execution. Through a detailed
evaluation, we observe that (a) provides an increase of up to 4.8% com-
pared to (b) and up to 11.5% compared to (c) in terms of average trip
execution, while all of them achieve close to the optimal performance.
At the same time, the optimal scales up to settings consisting of tenths
of EVs and a few hundreds of customer requests.

search, cooperative

## 1 Introduction

In a world where the majority of the population will be living in, or around,
large cities the current personal transportation model is not sustainable as it is
based almost entirely on privately owned internal combustion engine vehicles [1].

These vehicles cause high pollution (e.g., air and sound) and face low utilization rates [21]. Electric Vehicles (EVs) can be an efficient alternative to those using internal combustion engines in terms of running costs [8], environmental impact, and quality of driving. At the same time their main disadvantages are their short ranges and long charging times. To address such issues, cities usually resort to building a large number of charging stations. Such facilities are only worth building if there are enough EVs to use them. However, drivers will not buy EVs if charging stations are not first available, leading to a catch-22 situation.

In order to increase vehicle utilization, Mobility-on-Demand (MoD) schemes have been proposed [14]. MoD involves vehicles that are used by either individuals, or small groups of commuters, thus providing them with an alternative from using their privately owned vehicles. Such systems have the potential to reduce traffic congestion in urban areas, as well as the need for large numbers of parking spots and increase the vehicle utilization rates as few vehicles will cover the transportation needs of many commuters.

Given these benefits of EVs and MoD schemes, in this paper we explore scenarios within which EVs could be used within MoD schemes, and consider the related optimisation challenges. By addressing these challenges, the advantages of the two transportation modes can be combined [3], [14]. Moreover, the use of EVs in MoD schemes offers an opportunity to further market EVs to potential car owners as they get to try the technology before buying it. In this way, EV-equipped MoD schemes would help popularise EVs, while at the same time having a positive impact in urban traffic conditions as well as the environment.

Against this background, we model the MoD scheme for EVs and develop novel algorithms to solve the problem of scheduling trips for MoD consumers in order to maximize the number of trip requests serviced while coping with the limited range of EVs. We step upon the work presented in [17] and study the problem of assigning EVs to customers in a MoD scheme and we solve it offline and optimally using Mixed Integer Programming (MIP) techniques, as well as online using heuristic search. In doing so, we advance the state of the art as follows:

1. We extend the optimal scheduling algorithm "Off-Opt-Charge" presented in [17] which considers single travel requests by customers, by covering the option for customers to express their demand for more than one tasks, where as a task we consider a trip between a pair of locations starting a particular point in time.
2. We develop an online greedy scheduling algorithm for the problem of selecting the tasks to execute and the assignment of EVs to customers and we propose three alternative heuristic functions.

The rest of the paper is structured as follows: Section 2 presents related work, Section 3 formally defines the problem, Section 4 presents the optimal offline solution of the problem and Section 5 the equivalent online one. Section 6 provides a detailed evaluation of the algorithms and finally, Section 7 concludes and presents ideas for future work.

## 2   Related Work

In this context, Pavone et al. have developed mathematical programming-based rebalancing mechanisms for deciding on the relocation of vehicles to restore imbalances across a MoD network, either using robotic autonomous driving vehicles [16], or human drivers [15], while Smith et al. [19] use mathematical programming to optimally route such rebalancing drivers. Moreover, Carpenter et al. [4] propose solutions for the optimal sizing of shared vehicle pools. However, these works assume normal cars, while EVs present new challenges for MoD schemes. For example, EVs have a limited range that requires them to charge regularly. Moreover, if such MoD schemes are to become popular, it is important to ensure that charging capacity is managed and scheduled to allow for the maximum number of consumer requests to be serviced across a large geographical area. In addition, in order for MoD schemes to be economically sustainable, and given the higher cost of buying EVs compared to conventional vehicles, it is important to have them working at maximum capacity and servicing the maximum number of customers around the clock.

In such a setting, Drwal et al. [10] consider on-demand car rental systems for public transportation. To address the unbalanced demand across stations and maximise the operator's revenue, they adjust the prices between origin and destination stations depending on their current occupancy, probabilistic information about the customers' valuations and estimated relocation costs. Using real data from an existing on-demand mobility system in a French city, they show that their mechanisms achieve an up to 64% increase in revenue for the operator and at the same time up to 36% fewer relocations. In addition, Rigas et al. [17] use mathematical programming techniques and heuristic algorithms to schedule EVs in a MoD scheme taking into consideration the limited range of EVs and the need to charge their batteries. The goal of the system is to maximize serviced customers. Cepolina and Farina [5] study the use of single-sitter compact-sized EVs in a MoD scheme operating in a pedestrian zone. The vehicles are shared throughout the day by different users and one way trips are assumed. However, here the authors also assume open ended reservation to exist (i.e., the drop-off time is not fixed), thus adding one more dimension to the problem. Given this, they propose a methodology that uses a random search algorithm to optimize the fleet size and distribution to maximize the number of serviced customers. Moreover, Turan et al. [22] study the financial implications of smart charging in MoD schemes and they conclude that investing in larger battery capacities and operating more vehicles for rebalancing reduces the charging costs, but increases the fleet operational costs. Finally, Gkourtzounis et al. [12] propose a software package that allows for efficient management of a MoD scheme from the side of a company, and easy trip requests for customers.

From an algorithmic point of view, similarities can be found with problems such as the capacitated vehicle routing problem [6] which is a special case of the Vehicle Routing Problem [7], where each vehicle has a limited carrying capacity, the project scheduling problem [20], and the machine scheduling problem [13].

Overall, the need for battery charging as well as the strict order of task execution differentiate our problem compared to the majority of the works presented so far, and make it harder to find the optimal solution. Also the efficient online algorithms make it more applicable in real-world deployments. In the next section, the problem is formally defined.

## 3   Problem definition

In a MoD scheme which uses EVs, customers may choose to drive between pairs of predefined locations. They can choose at least one starting point and at least one end point. Since the MoD company's aim is to serve as many requests as possible, the system selects to execute the task which keeps the system in balance (i.e., trying to match demand across supply). A task is defined as a trip from a pick-up to drop-off location starting a particular point in time. Thus, based on the number of start and end points the customer has defined, all possible combinations are calculated and the equivalent tasks are created. We consider a network of parking stations where the EVs use to park and charge their batteries. The stations are considered as nodes aiming to be kept neither empty nor overloaded. The system needs to be in balance since the overloading of one station may cause major disruption to the network. A summary of all notations can be found in Table 1

We consider a fully connected directed graph $G(L, E)$ where $l \in L \subseteq \mathbb{N}$ is the set of locations where the stations exist and $e \in E \subseteq \mathbb{N}$ are the edges connecting all locations combinations. Each station has a maximum capacity $c_l^{max} \in \mathbb{N}$ declaring the number of EVs that can reside at it simultaneously. We assume a set of discrete time points $t \in T \subseteq \mathbb{N}$ where the time is global for the system and the same for all agents. We have a set of tasks $r \in R \subseteq \mathbb{N}$ where a task is a trip initiating a particular point in time. Thus, each task has a starting location $l_r^{start}$, an end location $l_r^{end}$, as well as a starting time $t_r^{start}$, a duration $\tau_r$ and an equivalent energy demand $e_r \in \mathbb{N}$.

We denote the set of EVs $a \in A \subseteq \mathbb{N}$. Each EV has a current location $l_{a,t} \in L$, a current battery level $e_{a,t} \in \mathbb{N}$, a maximum battery level $e_a^{max} \in \mathbb{N}$, an energy consumption rate $con_a \in \mathbb{N}$ where $con_a =$(energy unit/time point), a maximum travel time $\tau_a = e_a^{max}/con_a$ and a charging rate $ch_a \in \mathbb{N}$. Note that an EV changes location only when being driven by a customer and no relocation of vehicles exists.

Finally, we have a set of customers $i \in I \subseteq \mathbb{N}$ where a customer needs to travel between one or more pairs of locations $dem_i \subseteq R$. Customers act as fully cooperative agents when communicating their demand to the MoD company. After the demand is communicated to the company, an EV-to-customer assignment algorithm is applied. In doing so, a set of assumptions are made:

1. The MoD company is a monopoly. At this point competition between companies is not taken into consideration. This would introduce different approaches in decision making strategy and should include more variables into the problem domain (energy and labor cost, building rents, taxes, etc.), which are not the case in this paper.

2. The MoD company uses the same EV model. For simplification reasons it is considered that all EVs are of the same make and model.

3. All stations have one charger for each parking spot. This means that if there is a parking spot available, there is also a charger available. There is no waiting queue for charging.

4. EVs' full battery capacity is sufficient to make a journey from one station to any other without extra charge needed. No stops are required, and no charging time needs to be spent in between two locations. Travelling to locations beyond the maximum range of an EV needs a different formulation and induce challenges which will be solved in future work.

| Notation | Explanation |
|---|---|
| $l$ | Location of a station |
| $e$ | Edge connecting two stations |
| $c_l^{max}$ | Maximum capacity of a station |
| $t$ | Time point |
| $r$ | A task |
| $l_r^{start}$ | Start location of a task |
| $l_r^{end}$ | End location of a task |
| $t_r^{start}$ | Start time of a task |
| $\tau_r$ | Duration of a task |
| $e_r$ | Energy demand of a task |
| $a$ | An EV |
| $l_{a,t}$ | Current location of EV |
| $e_{a,t}$ | Energy level of EV |
| $e_a^{max}$ | Max battery capacity of EV |
| $ch_a$ | Charging rate of EV |
| $con_a$ | Consumption rate of EV |
| $\tau_a$ | Max travel time of EV |
| $i$ | A customer |
| $dem_i$ | Travel demand of customer |
| $\lambda_r$ | Task r accomplished (Boolean) |
| $\epsilon_{a,r,t}$ | True if EV a is working on task r at time t (Bollean) |
| $prk_{a,t,l}$ | True if EV a is parked at location l at time t (Boolean) |
| $bch_{a,t}$ | Charging rate of EV a at time t |

**Table 1.** Notations used in problem definition and algorithms.

## 4   Optimal Offline Scheduling

In this section, we assume that customer requests are collected in advance and we propose an optimal offline algorithm for the assignment of EVs to customers. The aim of this formulation is to maximize the number of tasks that are completed (a.k.a. customer service) (Equation 1). To achieve this, we present a solution based on Mixed Integer Programming (MIP) (solved using IBM ILOG CPLEX 12.10), where we use battery charging to cope with the EVs' limited range. MIP techniques have been particularly useful to solve such large combinatorial problems (e.g., combinatorial auctions [18], [2], travelling salesman problem [9]). We will refer to this algorithm as *Optimal*. In more detail, we define four decision variables: 1) $\lambda_r \in \{0, 1\}$ denoting whether a task $r$ is accomplished or not, 2) $\epsilon_{a,r,t} \in \{0, 1\}$ denoting whether EV $a$ is executing task $r$ at time $t$ or not, 3) $prk_{a,t,l} \in \{0, 1\}$ denoting whether EV $a$ is parked at time point $t$ at location $l$ or not and 4) $bch_{a,t} \in [0, ch_a]$ which denotes whether an EV $a$ is charging at time point $t$ and at which charging rate (i.e., the charging rate can be any between 0 and the maximum charging rate - $ch_a$).

**Objective function:**

$$max \sum_{r \in R} (\lambda_r) \tag{1}$$

**Subject to:**

– *Completion constraints:*

$$\sum_{a \in A} \sum_{t_r^{start} \leq t < t_r^{end}} \epsilon_{a,r,t} = \tau_r \times \lambda_r, \forall r \tag{2}$$

$$\sum_{a \in A} \sum_{t_r^{start} > t, t \geq t_r^{end}} \epsilon_{a,r,t} = 0, \forall r \tag{3}$$

$$\epsilon_{a,r,t+1} = \epsilon_{a,r,t} \forall a, \forall r, \forall t : t_r^{start} \leq t < t_r^{end} - 1 \tag{4}$$

$$bch_{a,t} \leq \sum_{l \in L} prk_{a,t,l} \times ch_a, \forall a, \forall t \tag{5}$$

$$0 \leq e_{a,t=0} + \sum_{t'=0}^{t} bch_{a,t'} - \sum_{r \in R} \sum_{t''=t_r^{start}}^{t} \epsilon_{r,a,t''} \times con_a \leq 100, \forall a, \forall t \tag{6}$$

$$\sum_{r \in dem_i} \lambda_r \leq 1, \forall i \tag{7}$$

The *completion* constraints ensure the proper execution of tasks. Thus, for each executed task, the time travelled must be equal to the duration of the trip concerned (Equation 2), while, at the same time no traveling must take place when a task is not executed (Equation 3). Moreover, each task is executed by only one EV at a time (Equation 4 together with Equation 9). Equation 5 ensures that each EV $a$ can charge only while being parked. When an EV is parked, it can charge with a charging rate up to its maximum one. However, when it is driving and $prk_{a,t,l} = 0$ it cannot charge. Regarding the time points the EV will charge and given the chosen formulation of the problem, the solver should choose any time points, as long as the available range will not compromise the task execution ability. At the same time, Equation 6 ensures that the battery level of an EV $a$ never exceeds 100% and never goes below 0%. Thus, no EV $a$ will execute a task $r$ for which it does not have enough range, nor will it charge more than its battery capacity. Note that we assume all EVs to have the same fixed average consumption. Finally, for each customer at most one of her alternative tasks $dem_i$ must be executed (Equation 7).

— *Temporal, spatial, and routing constraints:*

$$\sum_{l \in L} prk_{a,t,l} = 1 - \sum_{r \in R} \epsilon_{a,r,t}, \forall a, \forall t \tag{8}$$

$$2 \times \sum_{r \in R} \epsilon_{a,i,t_r^{start}} = \sum_{l \in L} \sum_{t \in T-1} |prk_{a,t+1,l} - prk_{a,t,l}|, \forall a \tag{9}$$

$$prk_{a,t_r^{start}-1,l_r^{start}} \geq \epsilon_{a,r,t_r^{start}}, \forall r, \forall a \tag{10}$$

$$prk_{a,t_r^{end},l_r^{end}} \geq \epsilon_{a,r,t_r^{end}}, \forall r, \forall a \tag{11}$$

$$\sum_{a \in A}(prk_{a,t,l}) \leq c_l^{max}, \forall l, \forall t \tag{12}$$

$$prk_{a,t=0,l} = l_a^{start}, \forall a, \forall l \tag{13}$$

$$\epsilon_{a,r,t=0} = 0, \forall a, \forall r \tag{14}$$

The *temporal, spatial and routing* constraints ensure the proper placement of the EVs over time. Equation 8 requires that for each time point at which an EV is executing a task, this EV cannot be parked at any location and also assures (together with Equation 4) that at each time point, each EV executes at most one task. Moreover, Equation 9 ensures that no EV changes location without executing a task (the sum of all changes of EVs' locations as denoted in $prk$ decision variable, must be double the total number of tasks that are executed). Note that, this constraint is linearized at run time by CPLEX.[3]

Now, whenever a task is to be executed, the EV that will execute this task must be at the task's starting location one time point before the task begins (Equation 10), and similarly, whenever a task has been executed, the EV that has executed this task must be at the task's end location the time point the task ends (Equation 11). Moreover, at every time point, the maximum capacity of each location must not be violated (Equation 12). Finally, at time point $t = 0$, all EVs must be at their initial locations (Equation 13), which also means that no tasks are executed at $t = 0$ (Equation 14).

---

[3] This is usually done by adding two extra decision variables and two extra constraints.

− *Cut constraints:*

$$\sum_{a \in A} prk_{a,t,l} = \sum_{a \in A} prk_{a,t-1,k} + \sum_{R^{start}(t,l)} \lambda_r - \sum_{R^{end}(t,l)} \lambda_r, \forall t, \forall l \qquad (15)$$

Equation 15 ensures that for every location, the total number of EVs at charging stations changes only when EVs depart or arrive to execute a task, or after executing tasks. Despite the fact that this constraint is covered by Equation 9, when added to the formulation, it significantly speeds up the execution time. In fact, it is known that the introduction of additional *cut constraints* into a MIP problem may cut off infeasible solutions at an early stage of the branch and bound searching process and thus reduce the time to solve the problem [11]. Given that MoD schemes should also work in a dynamic setting, in the next section we present an online greedy scheduling algorithm that uses alternatively three heuristic functions to solve the task execution problem.

## 5   Greedy Online Scheduling

In the previous section, we presented an optimal offline solution for the EV to customer assignment problem in a MoD setting. However, this algorithm assumes full knowledge of supply and demand in advance. In this section, in order to have a more complete set of tools to tackle the pre-defined problem we propose a greedy online algorithm that calculates an EV to task assignment in real time as requests arrive to the system. This algorithm applies a one-step look ahead heuristic search mechanism and achieves near optimal performance and scales to thousands of EVs and tasks.

Given that EVs change locations only when being driven by customers, the tasks that an EV will be able to execute in the future are directly related to the ones it has already executed in the past (i.e., the end location of one task will be the start location for the next one). In large settings, normally not all tasks can be executed. Thus the selection of the ones to execute is of great importance, since each decision can affect future task execution.

The proposed scheduling algorithm uses three heuristic functions in deciding on whether to execute a task or not. The first is the sum of squares of parked EVs at each station (see Eq. 16). The larger this number, the more imbalance for the system. In this case, we select to execute the task that will lead the EV to the location that minimizes this sum and causes the least imbalance to the system. For example if we consider two stations each having three parking spots, and three EVs. If all three EVs are parked in one station (when a task/request will be accomplished), the outcome would be: $3^2 + 0^2 = 9$. However, if two EVs were parked at one station and one at the other, the outcome would be: $2^2 + 1^2 = 5$. We refer to this heuristic as *Square*.

The second heuristic is the destination station capacity percentage (see Eq. 17). In this case, we divide the sum of the parked EVs at location $l$ by its total capacity and we select to execute the task that will lead an EV to the location with the highest current capacity (i.e., the lower number of existing EVs). This calculation is used to discover each location's capacity percentage separately and aims to move EVs to locations where the supply is low. We refer to this heuristic as *Destination*.

Finally, the third heuristic is a simple random choice of the task to execute. We refer to this heuristic as *Random*.

$$sq_t = \sum_{l \in L} (\sum_{a \in A} \epsilon_{a,t,l})^2, t \in T \qquad (16)$$

$$dcp_{l,t} = (\sum_{a \in A} \epsilon_{a,t,l})/c_l^{max} \qquad (17)$$

In what follows, we provide a step-by-step description of the greedy scheduling algorithm (see Alg. 1). Based on the online execution of the algorithm, if at time point $t$ a new customer $i$ arrives and expresses her demand, then the set $dem_i$ of all possible tasks is created (line 2). Then, we update the energy level for all EVs. EVs are assumed to charge their battery every time point they are parked unless the battery is fully charged (lines 3-8). Note that in contrast to the optimal algorithm, here the EVs charge with the maximum rate. For each task in $dem_i$, we check whether the end location of it has enough capacity to receive one vehicle. If this is true, then we search the set of EVs to find the ones that are parked at the starting location of the task and have enough energy to execute the task. If at least one such EV exists, then this task is added to the set $dem_i^*$ that contains the list of feasible tasks (lines 9-23). The next step is to calculate for each of the feasible tasks, the $score_r$ using one of the three heuristic functions (lines 24-27). These scores are later sorted on ascending order and the task with the lower score is selected to be executed (lines 27-28). Once the task has been selected, the EV is assigned to it and its location is updated accordingly (lines 29-39). In the next section we present a detailed evaluation of our algorithms.

## 6   Evaluation

In this section, we evaluate our algorithms on a number of settings in order to determine their ability to handle potentially large numbers of tasks, and EVs. To this end, we use real locations of pick-up and drop-off points owned by ZipCar[4] in Washington DC, USA which are available as open data,[5] while the distance and duration of all trips were calculated using Google maps. The evaluation of our algorithms is executed in two main parts:

- EXP1: The performance of the online and offline algorithms in terms of the average number of serviced customers (i.e., executed tasks).
- EXP2: The execution time and the scalability of the algorithms.

---

[4] https://www.zipcar.com/washington-dc.
[5] http://opendata.dc.gov/datasets/.

---

**Algorithm 1** EV-to-Customer assignment algorithm.

---

**Require:** $A, T, L, dem_i, \forall a, l, r : e_{a,t}, \epsilon_{a,t,l}, w_{a,t,r}, l_r^{start}, l_r^{end}, e_r, \tau_r$
1: {If a new customer $i$ arrives at time point $t$ then:}
2: Create $dem_i$ which consists of the combination of all start and end points defined by the customer.
3: **for all** $(a \in A)$ **do**
4:     **for all** $(t' \in T : t' < t)$ **do**
5:     {Update the energy level of each EV}
6:         $e_{a,t} = e_{a,t} + (\sum_{l \in L} \epsilon_{a,t',l}) \times ch_a - (\sum_{r \in R} w_{a,t',r}) \times con_a$
7:     **end for**
8: **end for**
9: **for all** $(r \in dem_i)$ **do**
10: {If the end location of the task has enough capacity for an incoming vehicle:}
11:     **if** $(c_{l_r^{end},t+\tau_r} < c_{l_r^{end},t+\tau_r})$ **then**
12:         $FoundEV \leftarrow False$
13:         $a = 0$
14:         **while** $(Found = False$ AND $a < |A|)$ **do**
15:         {Search the set of EVs until an EV with current location equal to the initial location of the task that has enough energy to execute the task is found.}
16:             **if** $(\epsilon_{a,t,l_r^{start}} = 1$ AND $e_{a,t} > e_r)$ **then**
17:                 $Found \leftarrow True$
18:             **end if**
19:             $a = a + 1$
20:         **end while**{Update the set of feasible tasks.}
21:         $(dem_i^* \leftarrow dem_i^* + r)$
22:     **end if**
23: **end for**
24: **for all** $(r \in dem_i^*)$ **do**
25:     Calculate $score$ for each task using one of the three heuristic functions.
26: **end for**
27: Sort $score$ in ascending order.
28: Select to execute task $r$ that minimizes the heuristic function.
29: **for all** $(t' \in T : t' \geq \tau_r)$ **do**
30: {Set the new location of the EV after the execution of the task.}
31:     $\epsilon_{a,t',l_r^{end}} = 1$
32: **end for**
33: **for all** $t' \in T : t' > t$ **do**
34:     $\epsilon_{a,t',l_r^{start}} = 0$
35: **end for**
36: **for all** $(t' \in T : t' \geq t$ AND $t' < t + \tau_r)$ **do**
37: {Set the EV to be working on the task for the equivalent time points. }
38:     $w_{a,t',r} = 1$
39: **end for**

---

All experiments were executed in the following setting: 1) One time point was selected to be equal to 15 mins, and in total 58 time points exist (i.e., equivalent to the execution of the MoD service from 7:00 to 18:00). 2) 8 locations exist and tasks can be formulated based on one of 56 possible trips (i.e., the trips are the combinations of the locations that form the MoD scheme. However, locations too close to each other were ignored) and trips as well as starting times were randomly selected. Each location has a maximum capacity $c_l^{max} = 10$. 3) Each customer $i$ has a demand $dem_i$ of up to three alternative tasks. 4) The energy consumption rate for each EV $a$ is selected to be $con_a = 10$ and the charging rate $ch_a = 25$. This means that for each time point that an EV is working the battery level is reduced by 10 units of energy, and for each time point an EV is charging the battery level is increased by up to 25 units of energy (fast battery charging): The average range of an EV is currently at around 150km. We assume an average speed of 40km/hour which means that an EV can drive for 3.75 hours. In our evaluation setting, one time point is equal to 15 minutes, and 3.75 hours equal to 15 time points. Thus, $con_a = 10\%$ of battery for each time point. A fast charger can fully charge an EV at around one hour. Thus, $ch_a = 25\%$ of the battery for each time point. Both $con_a$ and $ch_a$ are configuration parameters and can be selected by the user. All experiments were executed on a Windows PC with an Intel i7-4790K CPU and 16 GB of RAM running at 2400MHz.

## 6.1 EXP1: Customer service

Here we investigate the performance trade-off incurred by the online algorithms in terms of average customer service against the optimal offline one. Initially, we study a setting with 15 EVs and up to 70 customers. Note that each customer expresses her demand for up to 3 alternative tasks, with an average number of 2, so the average number of tasks is approximately double the number of customers. As we observe in Figure 1, all online algorithms are close to the optimal with the best being the Squared having a 94.2% efficiency in the worst case, then is the Destination with a 93.3% efficiency in the worst case and last is the Random with a 86.9% efficiency in the worst case.

Aiming to see how the number of EVs affects the performance of the online algorithms, we set up an experiment with 100 customers and up to 35 EVs. As we can observe in Figure 2 the overall image is similar to the previous case with the Squared being the best, the Destination second and the Random third. However, it is interesting to notice that when the number of EVs is low (5 EVs) or large (35 EVs) the performance deficit of the Destination and Random is smaller compared to the case where 20 EVs exist. This can be explained by the fact that when the number of EVs is low the heuristics, as they are connected to the number of EVs, cannot make a big difference, while when the number of EVs is high the problem becomes easier to solve.

Finally, in order to evaluate the performance of the online algorithms in larger settings, we set up an experiment with 100 EVs, 100 time points and up o 1200 customers. As we can observe in Figure 3, up to around 500 customers all three algorithms have a similar performance, but later the Squared and Destination have a better performance and for 1200 customer the Destination has a 95.4% efficiency compared to the Squared and the Random a 89.7% efficiency compared to the Squared.

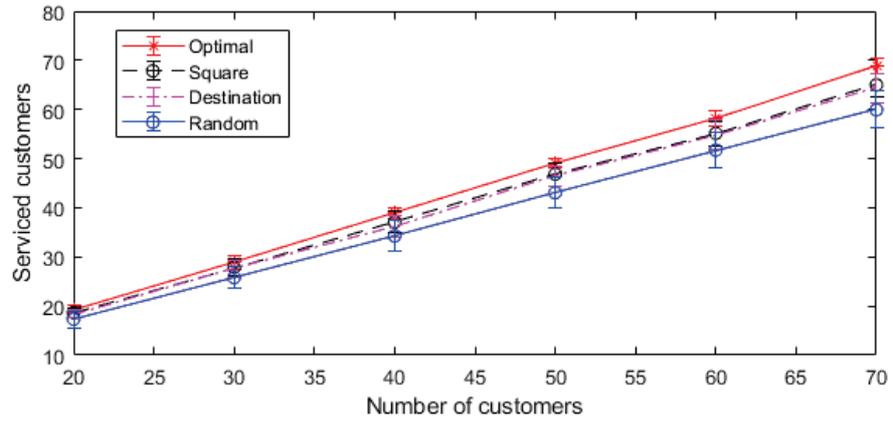

**Fig. 1.** Average number of serviced customers

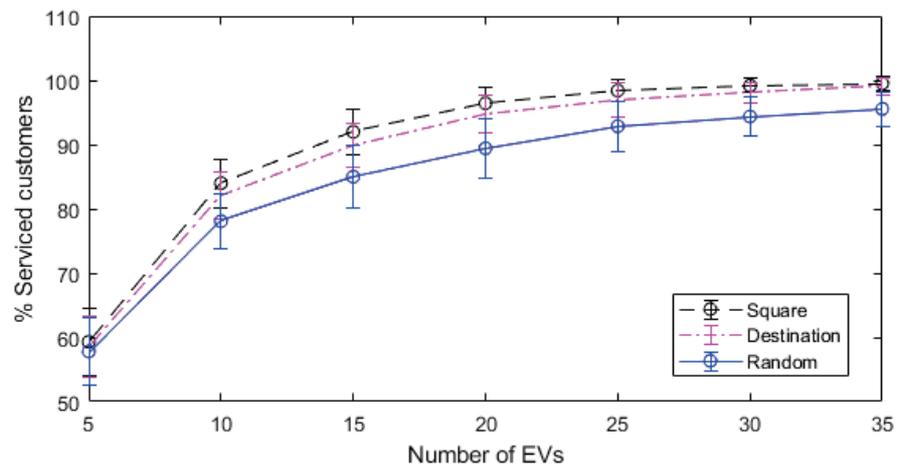

**Fig. 2.** Average number of serviced customers- Varying number of EVs

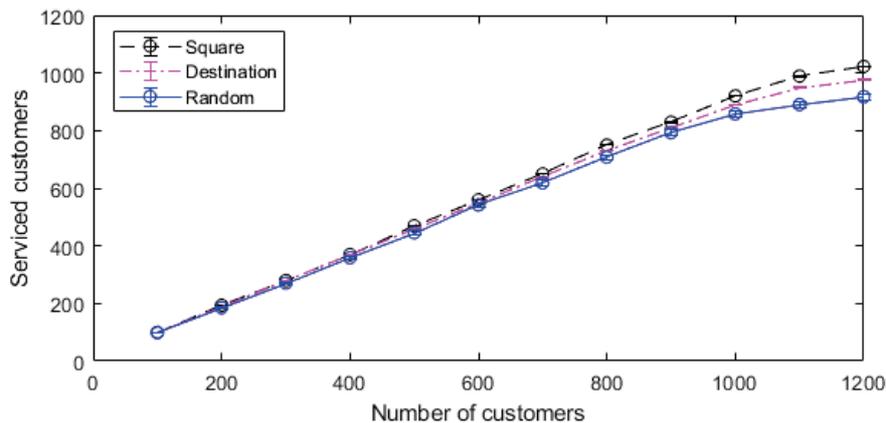

**Fig. 3.** Average number of serviced customers- Online algorithms

### 6.2   EXP2: Execution Time and Scalability

Execution time and scalability are typical metrics for scheduling algorithms. In a setting with 15 EVs and up to around 140 tasks (i.e., 70 customers), we see in Figure 4 that the execution time of the Optimal algorithm increases polynomially. Using MATLAB's Curve Fitting Toolbox we see that the Optimal's execution time is second degree polynomial with $R^2 = 97.11$. At the same time, the online algorithms have a very low execution time, as they all run in well under 0.05 seconds even in large settings.

## 7   Conclusions and future work

In this paper, we studied the problem of scheduling a set of shared EVs in a MoD scheme. We proposed an offline algorithm which collects the customers' demand in advance and calculates an optimal EV to customer assignment which maximizes the number of serviced customers. This algorithm scales up to medium sized problems. We also proposed three variations of an online algorithm which operates in a customer-by-customer basis and has shown to achieve near optimal performance while it can scale up to settings with thousands of EVs and locations.

Currently, we assume that the customer-agents are cooperative when communicating their demand to the system. As future work we aim to extend this by including non-cooperative agents and to apply mechanism design techniques in order to ensure truthful reporting. Moreover, we aim to improve the charging procedure of the EVs by trying to maximize the use of limited and intermittent energy from renewable sources. Finally, we want to enhance our algorithms in handling possible uncertainties in arrival and departure times, while aiming to maximize customer satisfaction and their profit.

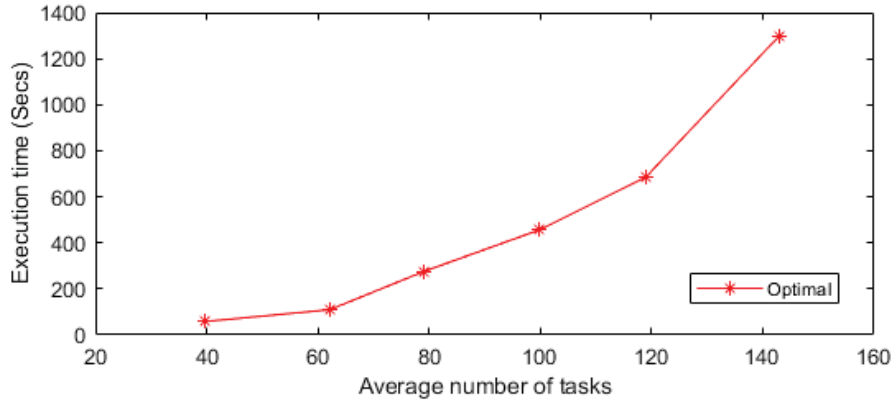

**Fig. 4.** Execution time of the Optimal algorithm

## Acknowledgment

This research is co-financed by Greece and the European Union (European Social Fund- ESF) through the Operational Programme "Human Resources Development,Education and Lifelong Learning" in the context of the project "Reinforcement of Postdoctoral Researchers - 2nd Cycle" (MIS-5033021), implemented by the State Scholarships Foundation (IKY).

## References

1. Administration, U.E.I.: Annual energy outlook 2019. Tech. rep. (2019)
2. Andersson, A., Tenhunen, M., Ygge, F.: Integer programming for combinatorial auction winner determination. In: MultiAgent Systems, 2000. Proceedings. Fourth International Conference on. pp. 39–46 (2000). https://doi.org/10.1109/ICMAS.2000.858429
3. Burns, L.D.: Sustainable mobility: a vision of our transport future. Nature **497**(7448), 181–182 (2013)
4. Carpenter, T., Keshav, S., Wong, J.: Sizing finite-population vehicle pools. IEEE Transactions on Intelligent Transportation Systems **15**(3), 1134–1144 (June 2014). https://doi.org/10.1109/TITS.2013.2293918
5. Cepolina, E.M., Farina, A.: A new shared vehicle system for urban areas. Transportation Research Part C: Emerging Technologies **21**(1), 230 – 243 (2012). https://doi.org/http://dx.doi.org/10.1016/j.trc.2011.10.005

6. Chandran, B., Raghavan, S.: Modeling and Solving the Capacitated Vehicle Routing Problem on Trees, pp. 239–261. Springer US, Boston, MA (2008)
7. Dantzig, G.B., Ramser, J.H.: The truck dispatching problem. Manage. Sci. **6**(1), 80–91 (Oct 1959)
8. Densing, M., Turton, H., Bäuml, G.: Conditions for the successful deployment of electric vehicles–a global energy system perspective. Energy **47**(1), 137–149 (2012)
9. Dorigo, M., Gambardella, L.M.: Ant colony system: a cooperative learning approach to the traveling salesman problem. IEEE Transactions on Evolutionary Computation **1**(1), 53–66 (Apr 1997). https://doi.org/10.1109/4235.585892
10. Drwal, M., Gerding, E., Stein, S., Hayakawa, K., Kitaoka, H.: Adaptive pricing mechanisms for on-demand mobility. In: Proceedings of the 16th Conference on Autonomous Agents and MultiAgent Systems. pp. 1017–1025. International Foundation for Autonomous Agents and Multiagent Systems (2017)
11. Floudas, C.A., Lin, X.: Mixed integer linear programming in process scheduling: Modeling, algorithms, and applications. Annals of Operations Research **139**(1), 131–162 (2005)
12. Gkourtzounis, I., Rigas, E.S., Bassiliades, N.: Towards online electric vehicle scheduling for mobility-on-demand schemes. In: European Conference on Multi-Agent Systems. pp. 94–108. Springer (2018)
13. Lomnicki, Z.A.: A "branch-and-bound" algorithm for the exact solution of the three-machine scheduling problem. Journal of the Operational Research Society **16**(1), 89–100 (1965). https://doi.org/10.1057/jors.1965.7, http://dx.doi.org/10.1057/jors.1965.7
14. Mitchel, W.J., Borroni-Bird, C.E., Burns, L.D.: Reinventing the automobile: Personal urban mobility for the 21st century. MIT Press (2010)
15. Pavone, M., Smith, S.L., Emilio, F., Rus, D.: Load balancing for mobility-on-demand systems. Robotics: Science and Systems (2011)
16. Pavone, M., Smith, S.L., Frazzoli, E., Rus, D.: Robotic load balancing for mobility-on-demand systems. The International Journal of Robotics Research **31**(7), 839–854 (2012). https://doi.org/10.1177/0278364912444766
17. Rigas, E.S., Ramchurn, S.D., Bassiliades, N.: Algorithms for electric vehicle scheduling in large-scale mobility-on-demand schemes. Artificial Intelligence **262**, 248 – 278 (2018). https://doi.org/https://doi.org/10.1016/j.artint.2018.06.006
18. Sandholm, T., Suri, S., Gilpin, A., Levine, D.: Winner determination in combinatorial auction generalizations. In: Proceedings of the First International Joint Conference on Autonomous Agents and Multiagent Systems: Part 1. pp. 69–76. AAMAS '02, ACM, New York, NY, USA (2002). https://doi.org/10.1145/544741.544760
19. Smith, S., Pavone, M., Schwager, M., Frazzoli, E., Rus, D.: Rebalancing the rebalancers: optimally routing vehicles and drivers in mobility-on-demand systems. In: American Control Conference (ACC), 2013. pp. 2362–2367 (June 2013). https://doi.org/10.1109/ACC.2013.6580187
20. Talbot, F.B., Patterson, J.H.: An efficient integer programming algorithm with network cuts for solving resource-constrained scheduling problems. Management Science **24**(11), 1163–1174 (1978)
21. Tomic, J., Kempton, W.: Using fleets of electric-drive vehicles for grid support. Journal of Power Sources **168**(2), 459 – 468 (2007). https://doi.org/http://dx.doi.org/10.1016/j.jpowsour.2007.03.010
22. Turan, B., Tucker, N., Alizadeh, M.: Smart charging benefits in autonomous mobility on demand systems. In: 2019 IEEE Intelligent Transportation Systems Conference (ITSC). pp. 461–466 (2019)